\documentclass[twoside,twocolumn,9pt]{article}
\usepackage{extsizes}
\usepackage[super,sort&compress,comma]{natbib} 
\usepackage[version=3]{mhchem}
\usepackage[left=1.5cm, right=1.5cm, top=1.785cm, bottom=2.0cm]{geometry}
\usepackage{balance,mathptmx,sectsty,graphicx,lastpage}
\usepackage[format=plain,justification=justified,singlelinecheck=false,font={stretch=1.125,small,sf},labelfont=bf,labelsep=space]{caption}
\usepackage{float}
\usepackage{fancyhdr}
\usepackage{fnpos}
\usepackage[english]{babel}
\addto{\captionsenglish}{%
  
}
\usepackage{array}
\usepackage{droidsans}
\usepackage{charter}
\usepackage[T1]{fontenc}
\usepackage[usenames,dvipsnames]{xcolor}
\usepackage{setspace}
\usepackage[compact]{titlesec}
\usepackage{hyperref}
\usepackage{gensymb}
\DeclareUnicodeCharacter{2212}{-}

\usepackage{epstopdf}

\definecolor{cream}{RGB}{222,217,201}

\begin{document}

\pagestyle{fancy}
\thispagestyle{plain}
\fancypagestyle{plain}{
\renewcommand{\headrulewidth}{0pt}
}

\makeFNbottom
\makeatletter
\renewcommand\LARGE{\@setfontsize\LARGE{15pt}{17}}
\renewcommand\Large{\@setfontsize\Large{12pt}{14}}
\renewcommand\large{\@setfontsize\large{10pt}{12}}
\renewcommand\footnotesize{\@setfontsize\footnotesize{7pt}{10}}
\makeatother

\renewcommand{\thefootnote}{\fnsymbol{footnote}}
\renewcommand\footnoterule{\vspace*{1pt}%
\color{cream}\hrule width 3.5in height 0.4pt \color{black}\vspace*{5pt}} 
\setcounter{secnumdepth}{5}

\makeatletter 
\renewcommand\@biblabel[1]{#1}            
\renewcommand\@makefntext[1]%
{\noindent\makebox[0pt][r]{\@thefnmark\,}#1}
\makeatother 
\renewcommand{\figurename}{\small{Fig.}~}
\sectionfont{\sffamily\Large}
\subsectionfont{\normalsize}
\subsubsectionfont{\bf}
\setstretch{1.125} 
\setlength{\skip\footins}{0.8cm}
\setlength{\footnotesep}{0.25cm}
\setlength{\jot}{10pt}
\titlespacing*{\section}{0pt}{4pt}{4pt}
\titlespacing*{\subsection}{0pt}{15pt}{1pt}

\fancyfoot{}
\fancyfoot[LO,RE]{\vspace{-7.1pt}\includegraphics[height=9pt]{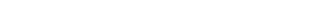}}
\fancyfoot[CO]{\vspace{-7.1pt}\hspace{13.2cm}\includegraphics{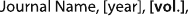}}
\fancyfoot[CE]{\vspace{-7.2pt}\hspace{-14.2cm}\includegraphics{head_foot/RF}}
\fancyfoot[RO]{\footnotesize{\sffamily{1--\pageref{LastPage} ~\textbar  \hspace{2pt}\thepage}}}
\fancyfoot[LE]{\footnotesize{\sffamily{\thepage~\textbar\hspace{3.45cm} 1--\pageref{LastPage}}}}
\fancyhead{}
\renewcommand{\headrulewidth}{0pt} 
\renewcommand{\footrulewidth}{0pt}
\setlength{\arrayrulewidth}{1pt}
\setlength{\columnsep}{6.5mm}
\setlength\bibsep{1pt}

\makeatletter 
\newlength{\figrulesep} 
\setlength{\figrulesep}{0.5\textfloatsep} 

\newcommand{\topfigrule}{\vspace*{-1pt}%
\noindent{\color{cream}\rule[-\figrulesep]{\columnwidth}{1.5pt}} }

\newcommand{\botfigrule}{\vspace*{-2pt}%
\noindent{\color{cream}\rule[\figrulesep]{\columnwidth}{1.5pt}} }

\newcommand{\dblfigrule}{\vspace*{-1pt}%
\noindent{\color{cream}\rule[-\figrulesep]{\textwidth}{1.5pt}} }

\makeatother

\twocolumn[
  \begin{@twocolumnfalse}
{\includegraphics[height=30pt]{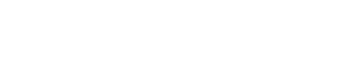}\hfill\raisebox{0pt}[0pt][0pt]{\includegraphics[height=55pt]{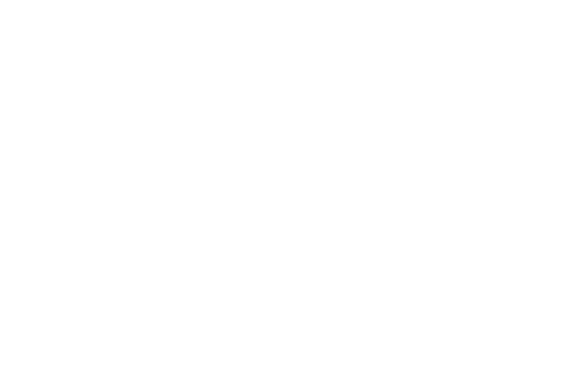}}\\[1ex]
\includegraphics[width=18.5cm]{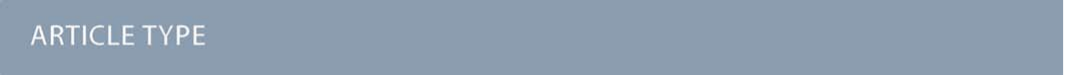}}\par
\vspace{1em}
\sffamily
\begin{tabular}{m{4.5cm} p{13.5cm} }

\includegraphics{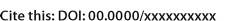} & \noindent\LARGE{\textbf{Unveiling mussel plaque core ductility: the role of pore distribution and hierarchical structure\dag~}} \\
\vspace{0.3cm} & \vspace{0.3cm} \\

 & \noindent\large{Yulan Lyu,\textit{$^{a,b}$} Mengting Tan,\textit{$^{c}$} Yong Pang,\textit{$^{b}$} Wei Sun,\textit{$^{a}$} Shuguang Li,\textit{$^{a}$} and Tao Liu,$^{\ast}$\textit{$^{b}$} } \\

\includegraphics{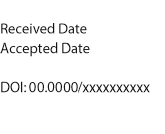} & \noindent\normalsize{The mussel thread-plaque system exhibits strong adhesion and high ductility, allowing it to adhere to various surfaces. While the microstructure of plaques has been thoroughly studied, the effect of their unique porous structure on ductility remains unclear. This study firstly investigated the porous structure of mussel plaque cores using scanning electron microscopy (SEM). Two-dimensional (2D) porous representative volume elements (RVEs) with scaled distribution parameters were generated, and the calibrated phase-field modelling method was applied to analyse the effect of the pore distribution and multi-scale porous structure on the failure mechanism of porous RVEs. The SEM analysis revealed that large-scale pores exhibited a lognormal size distribution and a uniform spatial distribution. Simulations showed that increasing the normalised mean radius value ($\bar{u}$) of the large-scale pore distribution can statistically lead to a decreasing trend in ductility, strength and strain energy, but cannot solely determine their values. The interaction between pores can lead to two different failure modes under the same pore distribution: progressive failure mode and sudden failure mode. Additionally, the hierarchical structure of multi-scale porous RVEs can further increase ductility by 40\%-60\% compared to single-scale porous RVEs by reducing stiffness, highlighting the hierarchical structure could be another key factor contributing to the high ductility. These findings deepen our understanding of how the pore distribution and multi-scale porous structure in mussel plaques contribute to their high ductility and affect other mechanical properties, providing valuable insights for the future design of highly ductile biomimetic materials.} \\

\end{tabular}

 \end{@twocolumnfalse} \vspace{0.6cm}

  ]

\renewcommand*\rmdefault{bch}\normalfont\upshape
\rmfamily
\section*{}
\vspace{-1cm}

\footnotetext{\textit{$^{a}$~Faculty of Engineering, University of Nottingham, Nottingham NG7 2RD, UK}}
\footnotetext{\textit{$^{b}$~School of Engineering and Materials Science, Queen Mary University of London, London E1 4NS, UK E-mail:tao.liu@qmul.ac.uk}}
\footnotetext{\textit{$^{c}$~School of Mechanical Engineering, Nanjing University of Science and Technology, Nanjing 210094, Jiangsu, China}}

\footnotetext{\dag~Electronic Supplementary Information (ESI) available: [details of any supplementary information available should be included here]. See DOI: 10.1039/cXsm00000x/}


\section{Introduction}
\label{introduction}
\begin{figure*}
 \centering
 \includegraphics[height=10cm]{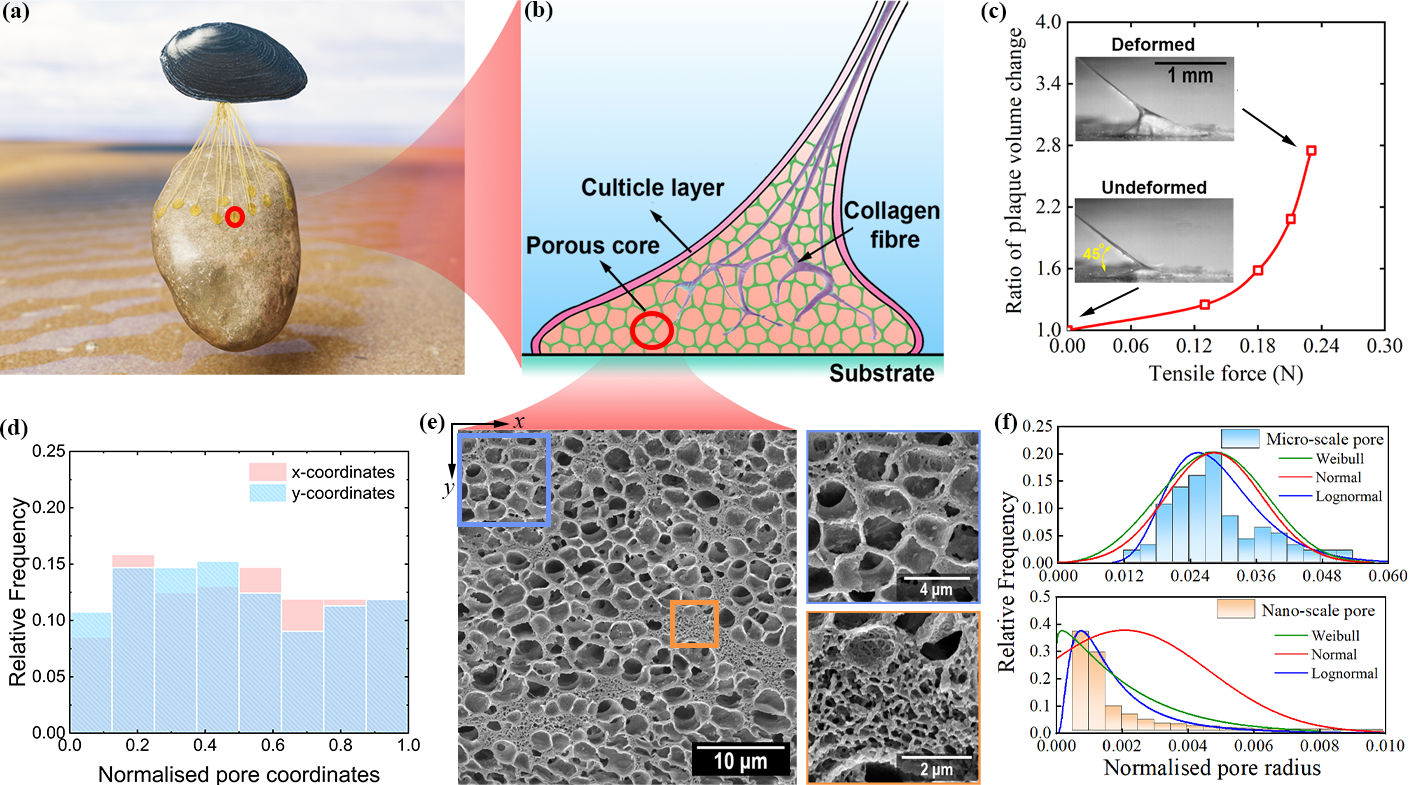}
 \caption{(a) Mussel byssus system attached on a rock, (b) schematic diagram of the mussel plaque’s microstructure \cite{pang2023quasi}, (c) large deformation of mussel plaque under tension, (d) the distribution of normalized x and y coordinates for large pore centres within the mussel plaque core, (e) the SEM photo showing the multi-scale porous structure of mussel plaque core and (f) the distribution of normalized radius for both large-scale (mirco) and small-scale (nano) pores within the mussel plaque core.}
 \label{Fig1}
\end{figure*}

The mussel byssus system, consisting of protein threads and porous plaques (Figs. \ref{Fig1}a and b), can anchor to almost any surface in harsh marine environments \cite{benedict1986composition, waite2022following, mcmeeking2019force, aldred2006mussel}. This ‘anchoring’ system enables mussels to withstand a force equivalent to approximately 10 times their own weight (Fig. \ref{Fig1}a) \cite{pang2023quasi}. Meanwhile, the high ductility of mussel plaques allows them to undergo significant deformation without fracturing under the strong hydrodynamic forces exerted by the ocean \cite{pang2024}. The plaque's volume can expand up to 2.8 times its original size when subjected to tension (Fig. \ref{Fig1}c). However, despite thorough studies on the adhesive performance of mussel plaques \cite{desmond2015dynamics, waite2017mussel, qureshi2022mussel, bandara2013marine, silverman2007understanding}, the cause of their high ductility remains unclear.    
 
Mussel plaques consist of a porous core and a protective cuticle layer (Fig. \ref{Fig1}b) \cite{tamarin1976structure, filippidi2015microscopic,martinez2015interfacial,wilhelm2017influence}. The cuticle layer, approximately 4 $\mu m$ thick, is covered by granules about 0.5 $\mu m$ in size \cite{pang2023quasi}. The plaque features randomly distributed nano- and micro-scale pores, with diameters approximately ranging from 50-800 nm to 1-3 $\mu m$, respectively (Fig. \ref{Fig1}d, e and f) \cite{filippidi2015microscopic,bernstein2020effects}. The formation of porous structures is closely related to seawater temperature and pH levels \cite{filippidi2015microscopic,bernstein2020effects}. The plaque core and cuticle layer work together to enhance ductility and establish a robust load-bearing capacity \cite{lyu2024determining,pang2023quasi,desmond2015dynamics}. However, it remains ambiguous how the unique porous structure (Fig. \ref{Fig1}e) affects plaque core's mechanical properties, particularly the high ductility.

Previous research has suggested that the pore arrangement within the plaque cores might influence the stress heterogeneity and thus have a significant impact on their ductility \cite{ghareeb2018role}. However, this study mainly applied ordered pores rather than the natural pore arrangement found in mussel plaques. It is therefore crucial to understand the role of the natural porous structure (Fig. \ref{Fig1}e) on the mechanical behaviours of plaque cores.

This study is motivated by the need to investigate the impact of the plaque core's porous structure on its mechanical performance, with a focus on ductility. We investigated the effects of pore distribution (Figs. \ref{Fig1}d and f) and multi-scale porous structure (Fig. \ref{Fig1}e) on the mechanical properties of plaque cores. The actual pore distribution within plaque cores was determined using scanning electron microscope (SEM) analysis. Scaled distribution parameters were then applied to create two-dimensional (2D) porous representative volume elements (RVEs) for subsequent numerical and experimental investigations of the porous structures. The validated phase field method, incorporating the Neo-Hookean hyperelastic model, was used in the simulations to investigate the effect of pore distribution and multi-scale structure on the failure mechanism of porous RVEs. By fully understanding how the pore distribution and multi-scale porous structure influence the mechanical performance, this study provides valuable insights for designing highly ductile metamaterials inspired by mussel plaques.

\section{SEM characterisation of mussel plaque core}
\subsection{SEM sample preparation }
Blue mussels (Mytilus edulis) were collected from the intertidal zone off the coast of Hunstanton, UK. Subsequently, they were tied to small acrylic plates and kept in a water tank with cold (8°C-10°C), oxygenated, and continuously circulated and filtered artificial seawater. Sea salt (Tropic Marin Pro-Reef Sea Salt, UK) was mixed with fresh water to replicate the salinity and pH levels seen in natural seawater (salinity = 33 ppt and pH = 8). The plaques were separated from the acrylic plates using a razor blade (Accu-Edge® Disposable Microtome Blades S35). The collected plaques were fixed in a solution containing 3.7\% formaldehyde and 2.5\% glutaraldehyde for approximately four hours \cite{filippidi2015microscopic, pang2023quasi}. After that, plaque samples were placed in Milli-Q water at 4°C for three days.

The plaques were then sectioned into 20 \(\mu m\) thick slices at a temperature of −20°C using a cryostat (Leica CM1850) and immediately immersed in Milli-Q water to remove the mounting medium \cite{filippidi2015microscopic}. The plaque slices underwent dehydration through sequential concentrations of ethanol and were subjected to two drying cycles of hexamethyldisilane (HDMS) \cite{filippidi2015microscopic}. The dehydrated samples were sputter-coated with a 10 nm thick iridium layer for 120 seconds using a Quorum Q150V Plus and imaged using a JEOL 7100F FEG-SEM with an accelerating voltage set at 5 kV.

\subsection{The pore distribution analysis for mussel plaque core}

\begin{figure*}
 \centering
 \includegraphics[height=22cm]{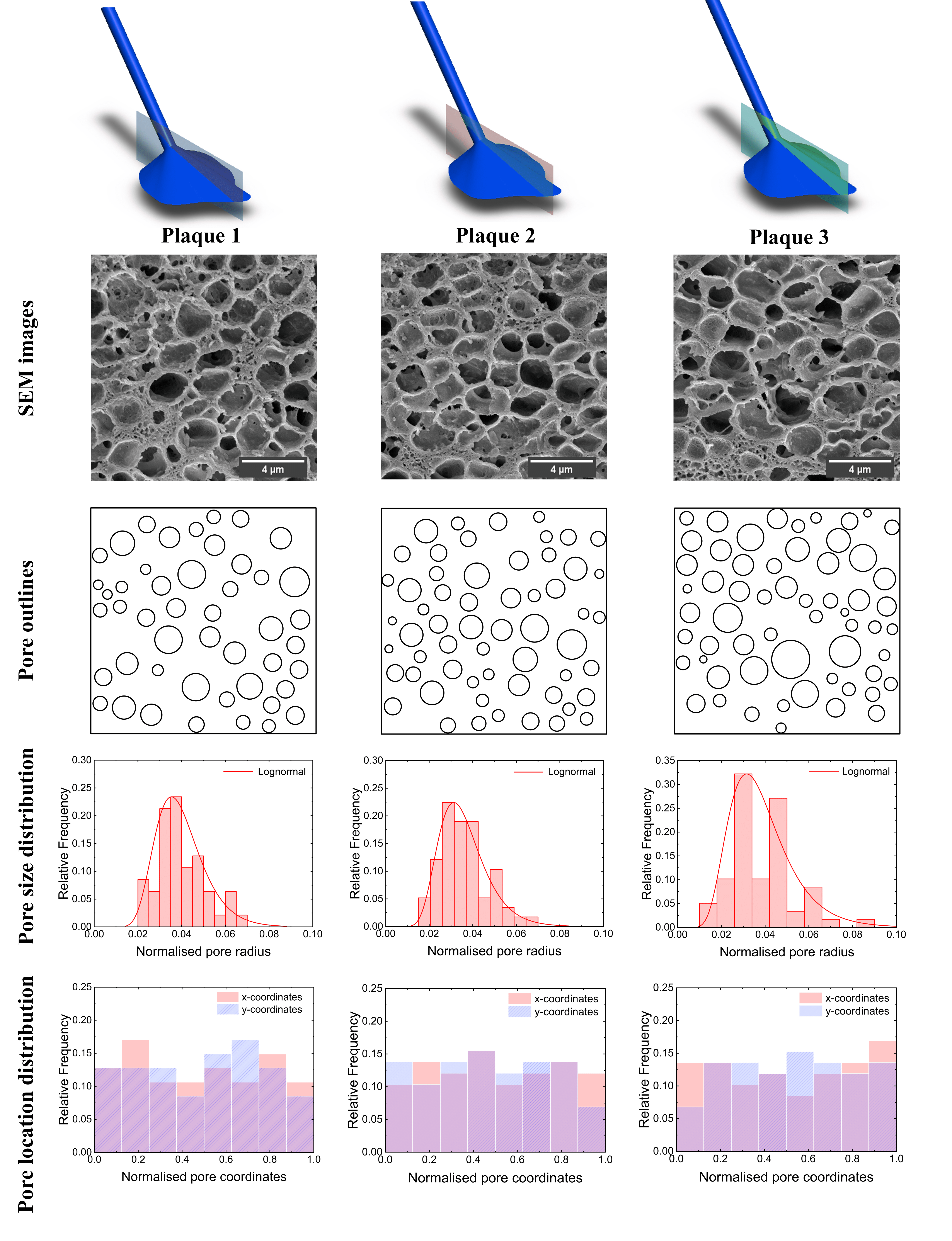}
 \caption{Pore size and location distribution analysis for SEM images of different plaques.}
 \label{Fig2}
\end{figure*}

Two-dimensional (2D) SEM images have suggested that the pores within the plaque cores are of two different scales, ranging in diameter from 50-800 nm (small-scale pores) to 1-3 $\mu m$ (large-scale pores) (Fig. \ref{Fig1}e). The pore distribution of large-scale pores within those SEM photos was first examined. In order to ensure the representativeness of the large-scale pore distribution in these photos, a total of nine SEM photos from different plaques or different locations of the same plaque were analysed (Figs. \ref{Fig1}e, \ref{Fig2}, S1 and S2 in the ESI,\dag~). The irregular large-scale pores were simplified into circular shapes, facilitating a more direct method for quantifying important distribution factors such as porosity, pore radius, and the coordinates of pore centres, which in turn ensures consistency of analysis. Despite the simplification to circular pores, the main geometric features of plaque cores were retained, thus ensuring that the pore distribution remained consistent with the original complex structure (Fig. S3 in the ESI,\dag~).

Meanwhile, the radii of large-scale pores and the coordinates of the pore centres were normalised by dividing the length of the squared SEM image edges (\(l_s\)) for the distribution analysis. The normalised mean value ($\bar{\mu}_p^*$) and normalised standard deviation value ($\bar{s}_p^*$) of the large-scale pores’ radius were calculated as: 
\begin{equation}
\bar{\mu}_p^*=\frac{r_{p 1}+r_{p 2}+\cdots+r_{p m}}{m * l_s}
\end{equation}

\begin{equation}
\bar{s}_p^*=\sqrt{\frac{\sum_{i=1}^m\left(\frac{r_{p i}}{l_S}-\bar{\mu}_p^*\right)^2}{m}}
\end{equation}

where $r_{p i}$ is the radius of large-scale pores within the SEM images and \(m\) is the number of the large pores within the SEM images. 

The porosity of large-scale pores within the selected SEM image can be calculated as: 
\begin{equation}
\upsilon_l=\frac{\pi r_{p 1}^2+\pi r_{p 2}^2+\cdots+\pi r_{p m}^2}{{l_s}^2}
\end{equation}

It can be observed that the large-scale pore radius consistently follows a lognormal distribution (Figs. \ref{Fig1}e, \ref{Fig2}, S1 and S2 in the ESI,\dag~). However, the distribution parameters and porosity may be affected by the position of the slices or plaques considered (see Tables \ref{tbl:example1}, S1 and S2 in the ESI,\dag~). The range of \(\bar{\mu}_p^*\) within the plaques is approximately 0.021-0.043, while the range of \(\bar{s}_p^*\) is approximately 0.005-0.014. The porosity of large-scale pores (\(\upsilon_l\)) ranges from 23\% to 33\%. Additionally, the Q-Q plots (quantile-quantile plots) (Fig. S4 in in the ESI,\dag~) comparing the x and y coordinates of large-scale pore centres in the nine SEM images (Figs. \ref{Fig1}e, \ref{Fig2}, S1 and S2 in the ESI,\dag~) show that both the x and y coordinates are consistent with a uniform distribution. 

\begin{table}[h]
\small
  \caption{\ Porosity (\(\upsilon_1\)), normalised mean radius (\(\bar{\mu}_p^*\)) and normalised standard deviation of radius (\(\bar{s}_p^*\)) of the large-scale pores within the SEM images for different plaques.}
  \label{tbl:example1}
  \begin{tabular*}{0.48\textwidth}{@{\extracolsep{\fill}}llll}
    \hline
    & Porosity (\(\upsilon_1\)) & \(\bar{\mu}_p^*\) & \(\bar{s}_p^*\)  \\
    \hline
    Plaque 1 & 25\% & 0.037 & 0.011  \\ 
    Plaque 2 & 26\% & 0.034 & 0.01  \\ 
    Plaque 3 & 30\% & 0.037 & 0.014  \\ 
    \hline
  \end{tabular*}
\end{table}

Therefore, it can be concluded that the radius of large-scale pores is lognormally distributed, and the coordinates of the large-scale pore centres are uniformly distributed within the mussel plaque core. Because the distribution parameters and porosity vary for each SEM photo, the parameters of Fig. \ref{Fig1}e were selected and scaled to generate the porous RVEs for further study: \(\upsilon_l\)=27\%, \(\mu_p^*\)=1.12 \(\mu m\) and \(s_p^*\)=0.34 \(\mu m\). \(\mu_p^*\) and \(s_p^*\) are the mean and standard deviation values of the lognormal distribution of the large-scale pores within Fig. \ref{Fig1}e, respectively. 

The pore radius distribution of the small-scale pores also follows a lognormal distribution with a mean value of 150 nm and a standard deviation value of 190 nm (Fig. \ref{Fig1}f). The porosity of the small-scale pores \(\upsilon_s\) in Fig. \ref{Fig1}e was calculated using ImageJ® as 5\%. The analysis for the small-scale pore distribution ensured that our main focus remained on the comprehensive study of large-scale pore distributions while still gaining relevant insights into the characteristics of small-scale pores. 

\section{The generation of the porous RVEs}
Two types of porous RVEs were generated for the numerical study. The first RVE type investigates the effect of large-scale pores, while the second RVE type examines the impact of the multi-scale architecture of the mussel plaque core.
\subsection{Generation of single-scale porous RVEs and post-processing procedure}

Test samples were manufactured by scaling the plaque pore sizes by a factor of 1000, transitioning from microns to millimetres. Two-dimensional (2D) square shaped porous RVEs were generated by using the scaled distribution parameters of the large-scale pores ($v_l$=27\%, $\mu_{p s}^*$=1.12 mm and $s_{p s}^*$=0.34 mm). The dimensions of the RVEs are 30 x 30 mm, comprising solely large-scale pores with the remaining volume filled with solid material.

The size effect of the RVEs was investigated by studying the responses of the RVEs with ordered pore arrangement and uniform pore size (details shown in the ESI,\dag~). The ordered porous RVEs featured porosity that was consistent with that of the large-scale pores obtained in the SEM image (\(v_l\)=27\%). The pore size within these RVEs was consistent with the scaled mean value of the large-scale pores ($\mu_{p s}^*$=1.12 mm), and the spacing between adjacent pores was the same (Fig. S5 in the ESI,\dag~). The results obtained from the simulations showed that the RVEs with the 30x30 mm dimension could reflect the macroscopic stress-strain relation of the porous structures (Fig. S6 in the ESI,\dag~). The macroscopic stress was determined by dividing the reaction force in the tensile direction by the cross-sectional area of the RVE, while the macroscopic strain was calculated by dividing the displacement in the tensile direction by the side length of the RVE. 

The commercial FE package ABAQUS was employed to create the single-scale RVEs using Python scripts. Within the Python scripts, the function \textit{random.lognormvariate (\(\mu_p\), \(s_p\))} was used to generate pores in the RVEs following a lognormal distribution, where \(\mu_p\) is the scaled mean value of the normal distribution corresponding to the lognormal distribution within the SEM image:
\begin{equation}
\mu_p=\ln \left(\frac{\mu_{p s}^{* 2}}{\sqrt{\mu_{p s}^{* 2}+s_{p s}^{* 2}}}\right)=0.07 \mathrm{~mm}
\end{equation}

\(s_p\) is scaled the standard deviation value of the normal distribution corresponding to the lognormal distribution within the SEM image:

\begin{equation}
s_p=\sqrt{\ln \left(\frac{s_{p s}^{* 2}}{\mu_{p s}^{* 2}}+1\right)}=0.3 \mathrm{~mm}
\end{equation}

The coordinates of pore centres follow a uniform distribution in accordance with the coordinate distribution observed in the SEM images (Figs. \ref{Fig1}d, \ref{Fig2}, S1 and S2 in the ESI,\dag~). The function \textit{random.uniform (low, high)} was applied in the python scripts to identify the locations of pores in the RVEs, where ‘low’ and ‘high’ are the min value and max value of pore centre coordinates, respectively. 

Therefore, in order to ensure all the pores are contained within the RVE, each pore’s centre coordinates (\(x_i\), \(y_i\)) must adhere to the following relation:
\begin{equation}
(x, y) \in\left\{\left(x_i, y_i\right) \mid r_i \leq x_i \leq w-r_i, r_i \leq y_i \leq w-r_i\right\}, i \in n
\end{equation}

here, \(n\) is the number of pores within the RVE, \(r_i\) represents the i th pore radius in the RVE, and \(w\) is the side length of the RVE, i.e. \(w\)= 30 mm. \(r_i\) and \(w-r_i\) present ‘low’ and ‘high’ in the function \textit{random.uniform(low, high)}, respectively.

Moreover, the distance between the centres of any two pores \(\theta\) in the RVE must be greater than the sum of their radii to avoid overlapping: 
\begin{equation}
\theta=\sqrt{\left(x_i-x_j\right)^2+\left(y_i-y_j\right)^2}>r_i+r_j, i \neq j, i, j \in n
\end{equation}

Consequently, the RVEs created through the ABAQUS Python scripts were required to fulfil all the previously mentioned criteria before being subjected to the FE analysis in ABAQUS.

The generated single-scale RVEs were discretised using 2D triangular plane stress elements, i.e.  CPS3T in ABAQUS notation and analysed using the phase field modelling method via analogy to a coupled heat transfer-displacement analysis. The phase field method is described in more detail in Section \ref{Phase_field} below. 

Following the completion of the ABAQUS calculation, a further Python script was developed with the objective of obtaining the following data from ABAQUS .odb files:

\begin{enumerate}
    \item The coordinates (\(x_i\), \(y_i\)) of all pore centres for each RVE \\ ($\left.i=1,2,3 \ldots n\right)$
    \item The radii $r_i$ of all pores for each RVE
    \item The number of pores for each RVE
    \item The macroscopic stress-strain curves for each RVE
\end{enumerate}

Furthermore, a MATLAB algorithm was created to extract and calculate various parameters from the stress-strain data sets. These include the strain energy density (\(\varphi\)), strength (\(\sigma_{max}\)), final failure strain (\(\varepsilon_{ff}\)), and the strain difference (\(\Delta\varepsilon\)) between the initial failure strain and the final failure strain for each RVE. In particular, the strain energy density (\(\varphi\)) was calculated as the area under the stress-strain curve. The initial failure strain was defined as the strain value corresponding to the peak strength, and the final failure strain (\(\varepsilon_{ff}\)) was the strain value corresponding to the point that stress decreased to 0. Additionally, the normalised mean value (\(\bar{\mu}\)) and normalised standard deviation value (\(\bar{s}\)) of the pores’ radius within each single-scale RVE were calculated as:

\begin{equation}
\bar{\mu}=\frac{r_1+r_2+\cdots+r_n}{n * w}
\end{equation}

\begin{equation}
\bar{s}=\sqrt{\frac{\sum_{i=1}^n\left(\frac{r_i}{w}-\bar{\mu}\right)^2}{n}}
\end{equation}

\subsection{Generation of multi-scale porous RVEs and post-processing procedure}
\begin{figure}[h]
\centering
  \includegraphics[height=3.5cm]{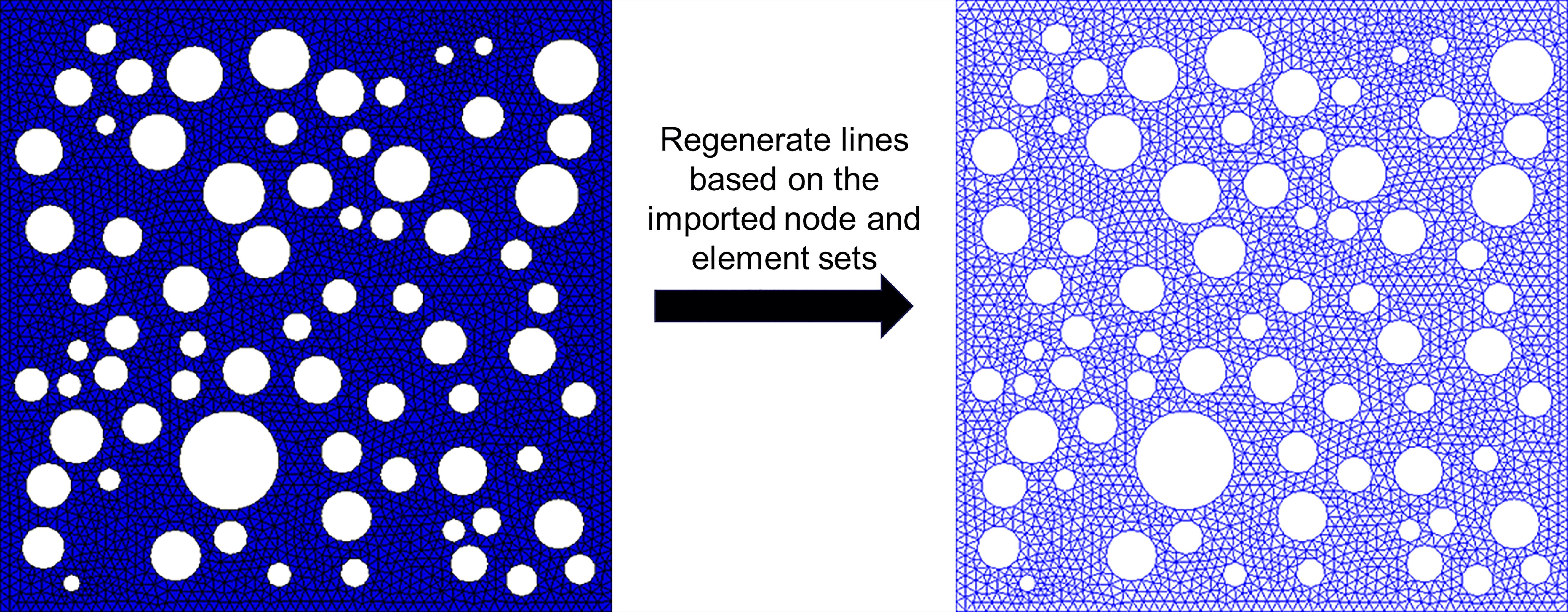}
  \caption{Schematic drawing of multi-scale porous RVE generation (The mesh element type switched from 2D triangular plane stress elements CPS3T to coupled temperature-displacement truss element T2D2T).}
  \label{Fig3}
\end{figure}

To further investigate the influence of multi-scale porous structures on mussel plaque behaviour, three different single-scale RVE FE models were selected. The multi-scale porous structures were generated via removing 2D elements in the FE models and connecting the nodes using truss elements of circular cross sections (i.e. coupled temperature-displacement truss element T2D2T) (Fig. \ref{Fig3}). The radius of the truss elements was determined to ensure that the newly generated RVEs reflected the same overall porosity (\(\upsilon=\upsilon_s+\upsilon_l=32\%\))  with the SEM image (Fig. \ref{Fig1}e). The approximate size of the small-scale pores within the multi-scale porous RVEs was set to 0.5 mm, which falls within the scaled range of small-scale pore sizes observed in mussel plaques (0.05 mm-0.8 mm). A size effect study, detailed in the ESI,\dag~, confirmed that the size of the small-scale pores within the multi-scale porous RVEs has an insignificant impact on the macroscopic material behaviours (Fig. S7 in the ESI,\dag~). Notably, the structural configuration adopted in the multi-scale porous RVEs serves as an approximation for examining the multi-scale porous structure of the mussel plaque. This simplified model was utilised to explore characteristic behaviours of such structures rather than a precise duplication of the plaque’s hierarchical structures.

Building on the same material properties and periodic conditions \cite{bhuwal2023discovery,Bhuwal2023,bhuwal2022data} used in the single-scale RVE FE models, the phase field method was applied to analyse the failure behaviour of the multi-scale porous RVEs (Section \ref{Phase_field}). The stress-strain curves of multi-scale porous RVEs were extracted from the ABAQUS .odb files, and fracture paths were analysed using the Abaqus/CAE graphical user interface. 

\section{ABAQUS implementation and experimental calibration of the phase field modelling method}
\label{Phase_field}
\subsection{Theory of phase field method for fracture modelling}
The phase field method was applied to analyse the failure mechanism of test samples and RVEs. The parent material was assumed to be an incompressible Neo-Hookean material. Therefore, in this section, a phase-field method for the incompressible Neo-Hookean material is introduced.

According to the theory of energy conservation, when a crack grows in an arbitrary body \(\Omega\), the energy per time unit $\dot{E}$ provided externally is equal to the rate of  the strain energy $\dot{U}$ stored in the material and the rate of the energy $\dot{D}^{d i s s}$ dissipated from the crack:
\begin{equation}
\label{5.10}
\dot{E}=\dot{U}+\dot{D}^{\text {diss}}
\end{equation}

Let \(c\) denote the phase field variable. The value of \(c\) is between 0 and 1, with \(c=0\) and \(c=1\) denoting undamaged materials and fully damaged materials, respectively. The quadratic degradation function \(g(c)\) is introduced to control the material stiffness, i.e.,
\begin{equation}
\label{5.11}
g(c)=(1-c)^2
\end{equation}

Therefore, the strain energy stored in the material can be written as the following:
\begin{equation}
U=\int_{\Omega} g(c) \varphi d V
\label{5.12}
\end{equation}
where \(\varphi\) is the strain energy density.

For incompressible 2D plane stress studies in this research, the strain energy density \(\varphi\) can be expressed as \cite{loew2019rate}:
\begin{equation}
\varphi=\varphi_{i s o}(\mathbf{F})
\label{5.13}
\end{equation}
where \(\mathbf{F}\) is deformation gradient tensor, \(\varphi_{i s o}(\mathbf{F})\) is the isochoric strain energy density.

Therefore, based on the vector calculus identity  \cite{bonet1997nonlinear} and divergence theorem \cite{pfeffer1986divergence}, the rate of the stored strain energy is given by:
\begin{equation}
\begin{aligned}
\dot{U}=&\int_{\Omega}\left(\frac{\partial g(c)}{\partial c} \varphi \dot{c}\right) d V+\int_{\partial \Omega} g(c) \frac{\partial \varphi}{\partial \mathbf{F}} \cdot \dot{\mathbf{u}} \cdot \mathbf{n} d A\\& -  \int_{\Omega}\left(\nabla \cdot\left(g(c) \frac{\partial \varphi}{\partial \mathbf{F}}\right)\right) \cdot \dot{\mathbf{u}} d V
\label{5.14}
\end{aligned}
\end{equation}
where \(\dot{c}\) is the rate of the damage variable, $\dot{\mathbf{u}}$ is the velocity field presenting the rate of change of the displacement field $\mathbf{u}$ with respect to time, $\partial \Omega$ is the surface of the volumetric body $\Omega$ and \(\mathbf{n}\) is the outward normal to the surface $\partial \Omega$. The derivation process for eqn (\ref{5.14}) is shown in the ESI,\dag~.

Meanwhile, the dissipated crack formation energy density \(\tau\) was given by Miehe et al. (2010) \cite{miehe2010phase}:
\begin{equation}
\tau=G_c \gamma_l(c, \nabla c)=G_c\left(\frac{c^2}{2 l}+\frac{l}{2}(\nabla c \cdot \nabla c)\right)
\end{equation}
where \(G_c\) is the critical energy release rate and \(\gamma_l(c, \nabla c)\) is the crack density function with length scale parameter \(l\) , phase field variable \(c\) and the spatial gradient \(\nabla c\) \cite{miehe2010phase}.

Therefore, the energy dissipated by the crack formation \(D^f\) and the rate of the dissipated energy from crack formation \(\dot{D}^f\) can be written as:
\begin{equation}
D^f=\int_{\Omega} \tau d V=\int_{\Omega} G_c\left(\frac{c^2}{2 l}+\frac{l}{2}(\nabla c \cdot \nabla c)\right) d V
\end{equation}
\begin{equation}
\label{5.17}
\dot{D}^f=\int_{\Omega} G_c\left(\frac{c \dot{c}}{l}+l(\nabla c \cdot \nabla \dot{c})\right) d V
\end{equation}

Again, based on the vector calculus identity \cite{bonet1997nonlinear} and divergence theorem \cite{pfeffer1986divergence}, eqn (\ref{5.17}) can be rewritten as:
\begin{equation}
\label{5.18}
\dot{D}^f=\int_{\Omega} G_c \dot{c}\left(\frac{c}{l}-l \nabla^2 c\right) d V+\int_{\partial \Omega} G_c l \nabla c \dot{c} \cdot \mathbf{n} d A
\end{equation}
The derivation process for eqn (\ref{5.18}) is in the ESI,\dag~.

Moreover, a rate of dissipation energy for crack growth \(\dot{D}^g\) was introduced by Pascal et al. (2019) \cite{loew2019rate}:
\begin{equation}
\dot{D}^g=\int_{\Omega} \kappa_1 \dot{c}^2 d V
\end{equation}
where scalar \(\kappa_1\) denotes a viscosity parameter. 
Thus, the rate of the dissipated energy through the crack can be presented as:
\begin{equation}
\begin{aligned}
\dot{D}^{\text {diss }}&=\dot{D}^f+\dot{D}^g\\&=\int_{\Omega} G_c \dot{c}\left(\frac{c}{l}-l \nabla^2 c\right) d V+\int_{\partial \Omega} G_c l \nabla c \dot{c} \cdot \mathbf{n} d A+\int_{\Omega} \kappa_1 \dot{c}^2 d V
\label{5.20}
\end{aligned}
\end{equation}

Additionally, the external energy can be broadly classified into work done by surface tractions (forces acting on the boundary of the body) and body forces (forces acting throughout the volume of the body). Hence, the rate of external energy \(\dot{E}\) can be expressed as:
\begin{equation}
\dot{E}=\int_{\Omega} \boldsymbol{b} \cdot \dot{\mathbf{u}} d V+\int_{\partial \Omega} \boldsymbol{t} \cdot \dot{\mathbf{u}} d A
\label{5.21}
\end{equation}
where \(\boldsymbol{b}\) presents the volumetric body force vector and \(\boldsymbol{t}\) is the surface traction force vector.

Therefore, based on eqns (\ref{5.10}), (\ref{5.14}), (\ref{5.20}) and (\ref{5.21}), it can be obtained as follows:
\begin{equation}
\begin{aligned}
\int_{\Omega} \boldsymbol{b} \cdot \dot{\mathbf{u}} d V+\int_{\partial \Omega} \boldsymbol{t} \cdot \dot{\mathbf{u}} d A= & \int_{\Omega} \dot{c}\left(\frac{\partial g(c)}{\partial c} \varphi+\frac{G_c c}{l}-G_c l \nabla^2 c+\kappa_1 \dot{c}\right) d V \\
& -\int_{\Omega}\left(\nabla \cdot\left(g(c) \frac{\partial \varphi}{\partial \boldsymbol{F}}\right)\right) \cdot \dot{\mathbf{u}} d V \\
& +\int_{\partial \Omega} g(c) \frac{\partial \varphi}{\partial \boldsymbol{F}} \cdot \dot{\mathbf{u}} \cdot \mathbf{n} d A +\int_{\partial \Omega} G_c l \nabla c \dot{c} \cdot \mathbf{n} d A
\end{aligned}
\end{equation}

Given the macroforce and microforce balances, we can finally obtain the following equations:
\begin{equation}
\nabla \cdot\left(g(c) \frac{\partial \varphi}{\partial F}\right)+\boldsymbol{b}=0
\label{5.23}
\end{equation}
\begin{equation}
\frac{\partial g(c)}{\partial c} \varphi+\frac{G_c c}{l}-G_c l \nabla^2 c+\kappa_1 \dot{c}=0
\label{5.24}
\end{equation}
with the following Neumann boundary conditions that may be applied 
\begin{equation}
g(c) \frac{\partial \varphi}{\partial F} \cdot \mathbf{n}=\mathbf{t} \text { and } G_c l \nabla c \cdot \mathbf{n}=\mathbf{0}
\label{5.25}
\end{equation}
Based on eqn (\ref{5.24}), if \(\kappa_1=0\) , the phase-field response is rate-independent. Alternatively, the response is rate-dependent.

\subsection{Finite element (FE) analysis implementation}
\label{FE}
\subsubsection{Heat transfer analogy.~~}
The strain energy density for the incompressible Neo-Hookean hyperelastic material specifically is shown below:
\begin{equation}
\varphi=C_1\left(I_1-3\right)
\label{5.26}
\end{equation}
where \(C_1\) is a material constant and \(I_1\) is the first invariant.

Combing eqns (\ref{5.11}), (\ref{5.24}) and (\ref{5.26}),
\begin{equation}
\frac{\kappa_1}{G_c l} \dot{c}-\nabla^2 c=\frac{2(1-c) \cdot\left[C_1\left(I_1-3\right)\right]}{G_c l}-\frac{c}{l^2}
\label{5.27}
\end{equation}

The time derivative term of phase field variable \(\dot{c}\) can be expressed as:
\begin{equation}
\dot{c}=\frac{d c}{d t}=\frac{\partial c}{\partial t}+\frac{\partial c}{\partial x} \mathrm{v}_x+\frac{\partial c}{\partial y} \mathrm{v}_y+\frac{\partial c}{\partial z} \mathrm{v}_z
\label{5.28}
\end{equation}
where \(x, y, z\) are the spatial coordinates; and \(\mathrm{v}_x, \mathrm{v}_y, \mathrm{v}_z\) are the corresponding velocity components.

Because the focus of our research is the progressive and slow fracture of 2D porous structures, it was assumed $\mathrm{v}_x \approx \mathrm{v}_y \approx \mathrm{v}_z \approx 0$. Therefore, eqn (\ref{5.28}) can be simplified as: 
\begin{equation}
\dot{c}=\frac{d c}{d t} \approx \frac{\partial c}{\partial t}
\label{5.29}
\end{equation}

Furthermore, eqn (\ref{5.27}) can be rewritten as: 
\begin{equation}
\frac{\kappa_1}{G_c l} \frac{\partial c}{\partial t}-\nabla^2 c=\frac{2(1-c) \cdot\left[C_1\left(I_1-3\right)\right]}{G_c l}-\frac{c}{l^2}
\label{5.30}
\end{equation}

Meanwhile, the field equation of heat transfer given by the first law of thermodynamics reads as,
\begin{equation}
\rho c_p \frac{\partial T}{\partial t}-k \nabla^2 T=q
\label{5.31}
\end{equation}
where \(T\) is the temperature field, \(\frac{\partial T}{\partial t}\) is the rate of the temperature field, \(k\) is the thermal conductivity, \(c_p\) is the specific heat, \(\rho\) is the density, and \(q\) is the heat source. 

Given the similarity of eqns (\ref{5.30}) and (\ref{5.31}), the phase field model can be implemented via the heat transfer analogy by letting the temperature variable \(T\) represent the phase field variable \(c\). Assuming \(k=1\), the equation for the heat source \(q\) can be expressed as:
\begin{equation}
q=\rho c_p \frac{\partial T}{\partial t}-\nabla^2 T \equiv \eta \frac{\partial c}{\partial t}-\nabla^2 c=\frac{2(1-c) \cdot\left[C_1\left(I_1-3\right)\right]}{G_c l}-\frac{c}{l^2}
\end{equation}
where \(\rho c_p\) is analogised as the fracture parameter \(\eta\), $\eta=\frac{\kappa_1}{G_c l}$.

\subsubsection{ABAQUS implementation.~~}
The phase field method was numerically implemented in ABAQUS using combined User subroutines UHYPER and HETVAL \cite{navidtehrani2021simple,abaqus2011abaqus}. The strain energy density \(\varphi\) of the Neo-Hookean hyperelastic model was degraded in UHYPER using eqn (\ref{5.11}). A local history variable field \(H\)  introduced by Miehe et al. (2010) \cite{miehe2010phase} was applied to store the value of the strain energy density within the UHYPER and transfer it to the HETVAL subroutine to achieve the temperature analogy. The history variable field H could prevent reversible damage and ensure that the damage variable continues to grow until completely damaged \cite{miehe2010phase}:
\begin{equation}
H=\max _{\beta \in[0, t]}\left(\varphi_{t=\beta}\right)
\end{equation}

Meanwhile, the heat flux (\(q\)) and the changing rate of the heat flux (\(\chi\)) were defined in the HETVAL subroutine as follows:
\begin{equation}
q=\frac{2(1-c)}{G_c l} H-\frac{c}{l^2}
\end{equation}
\begin{equation}
\chi=\frac{\partial q}{\partial T} \equiv \frac{\partial q}{\partial c}=-\left(\frac{1}{l^2}+\frac{2 H}{G_c l}\right)
\end{equation}

The coupled temp-displacement solver within the ABAQUS package was used for the FE calculations. The ‘steady-state’ solver option was employed for rate-independent phase-field analysis (i.e., \(\kappa_1 =0\)), while the ‘transient’ response was used for the rate-dependent phase field model. It is also necessary to define the thermal conductivity (\(k\)), density (\(\rho\)) and specific heat (\(c_p\)) within the material properties for the rate-dependent model, while only the thermal conductivity (\(k\)) needs to be defined for the rate-independent model. The material constant \(C_1\), the critical energy release rate \(G_c\) and the length scale \(l\) were needed to be defined for both models within the Hyperelastic 'User-defined' section. Additionally, the ‘Depvar’ and ‘Heat Generation’ options were active to achieve element deletion and the temperature field analogy, respectively. The temperature field was created inside the FE model with magnitude = 0 and the results of the nodal solution temperature (NT11) were plotted to visualise the phase field solution (Element with \(NT11\geq1\) was deleted). 

\subsection{Material parameters acquisition and model calibration}
\subsubsection{Single-edge notched tension (SENT) tests.~~}
\label{SENT}
Obtaining precise material properties of mussel plaque soft cores is challenging due to current limitations in technical and experimental methodologies. As the focus of this paper is to understand the effect of the microstructure rather than the material properties of the parent material, TangoBlack Plus FLX980, a soft and ductile 3D printing material was employed to reproduce the scaled microstructure in both numerical and experimental studies.

Single-edge notched tension (SENT) tests with different notch lengths were carried out to obtain the material parameters of TangoBlack Plus FLX980, including the material constant \(C_1\), critical energy release rate \(G_c\), and length scale \(l\). Rectangular strips using in the SENT tests had a total length of 100 mm (clamp distance 80 mm), a width of 25 mm, a thickness of 1 mm, a notch width of 0.3 mm, and variable notch lengths (\(\alpha\)) ranging from 0 mm to 4 mm (Fig. \ref{Fig4}a).\cite{roucou2019critical, rivlin1953rupture} The sample size was designed based on the rubber fracture experiments conducted by Rivlin and Thomas (1953) \cite{rivlin1953rupture} and Roucou et al. (2019) \cite{roucou2019critical}. Samples were 3D printed using TangoBlack Plus FLX980. SENT tests were displacement-controlled with a fixed strain rate of 100\% strain/min using an INSTRON tensile machine. Since the primary focus of this study does not involve the parent material, the effect of global strain rate on the parent material was not discussed. Instead, we ensured that the global strain rate was consistent for both SENT tests and following porous sample tensile tests to exclude the effects from the parent material. A Thorlabs DCC1545M CMOS camera, equipped with an imaging lens (focal length of 100 mm), was employed to monitor specimen deformation \cite{Pang20242}. The CMOS camera was configured at a frame rate of 10 fps and a special resolution of 9.6 pixel/mm.   

\(C_1\) and \(G_c\) were determined by analysing the nominal stress-strain curves obtained from SENT tests. \(C_1\) was derived by fitting the tensile stress-strain curves prior to failure, and the critical strain energy release rate \(G_c\) was computed using the equation below \cite{roucou2019critical, rivlin1953rupture}:
\begin{equation}
\begin{gathered}
G(\alpha, \lambda)=2 \alpha K(\lambda) W(\lambda) \\
G_c=G\left(\lambda_c\right)
\label{5.36}
\end{gathered}
\end{equation}
where \(G\) is the strain energy rate, \(\alpha\) is the initial notch length, \(W(\lambda)\) is the material strain energy density under uniaxial tension, \(\lambda_c\) is the critical stretch value obtained when the applied load abruptly drops to zero in the SENT test, and  \(K(\lambda)\) is a function of the stretch value \(\lambda\) that can be defined as \cite{roucou2019critical}:
\begin{equation}
K \simeq \frac{3}{\sqrt{\lambda}}
\label{5.37}
\end{equation}
SENT tests with notch sizes of 2 mm, 3 mm, and 4 mm were simulated to calibrate with experimental results and to obtain the value of the length scale (\(l\)). The simulation was carried out using the rate-independent phase field method outlined in Section \ref{Phase_field}, selected specifically for its suitability in modelling the slow fracture process ($\kappa_1 \approx 0$).

\subsubsection{Porous samples tensile tests.~~}
\begin{figure}[h]
\centering
  \includegraphics[height=6cm]{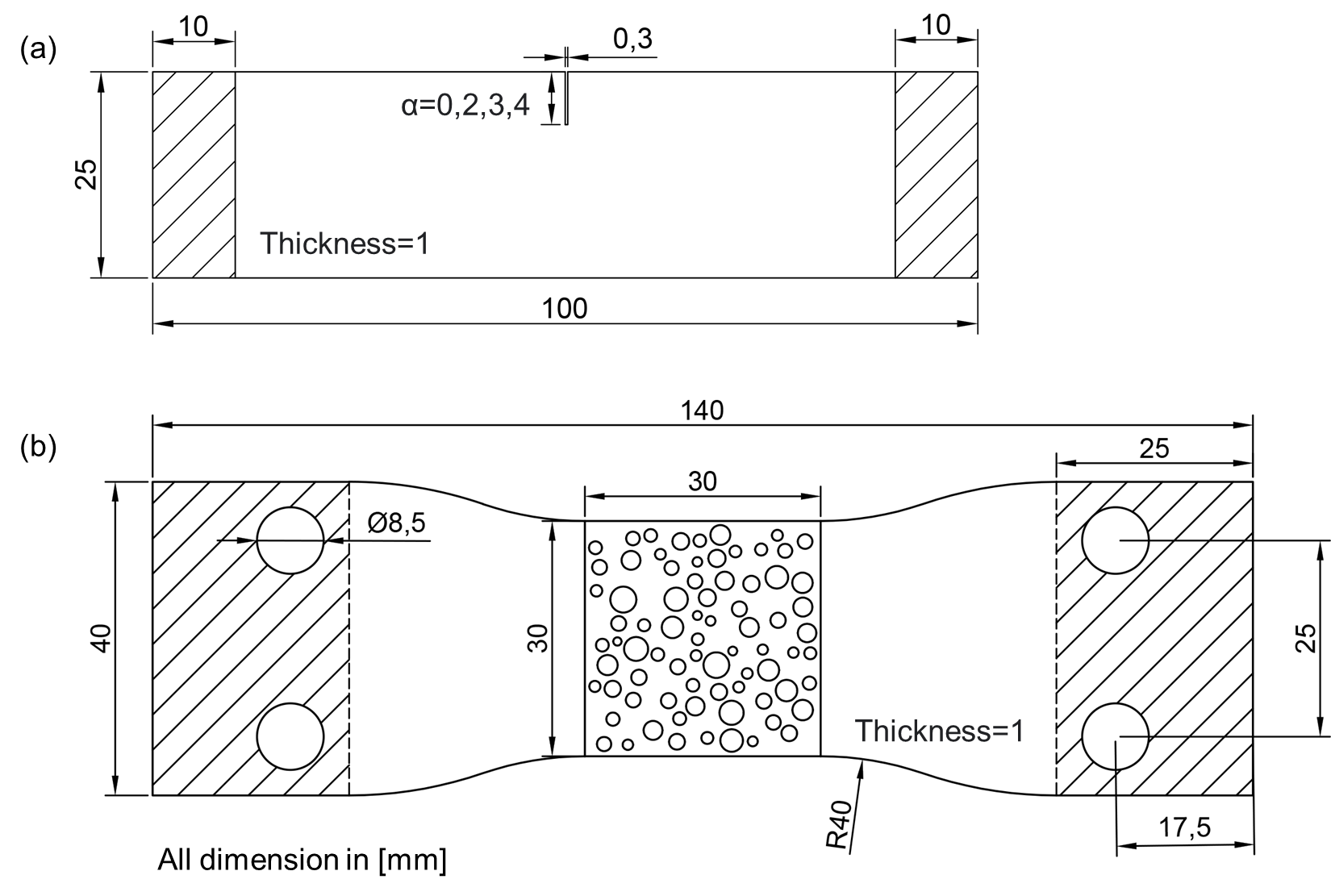}
  \caption{Schematic drawings of (a) the SENT test samples and (b) the porous samples for tensile testing.}
  \label{Fig4}
\end{figure}

To further validate the phase field method, three porous samples were 3D printed for uniaxial tensile testing. The porous sample design was based on the size of RVEs and used the scaled distribution parameters for large pores (Fig. \ref{Fig4}b). The uniaxial tensile tests were conducted at a fixed strain rate of 100\% strain/min, consistent with the conditions of the SENT experiments. Even though the global strain rate was the same, the local fracture of porous structure was a dynamic and fast process. Therefore, the phase field rate-dependent model was employed for the simulation. The experimental and simulated tensile stress-strain curves and crack paths for each sample were compared to calibrate the fracture parameter \(\eta\). Other material parameters, including the material constant (\(C_1\)), critical energy release rate (\(G_c\)) and length scale (\(l\)) were consistent with those calculated in Section \ref{SENT}. 

\subsubsection{Calibration results for the phase field modelling methods.~~}
\begin{figure}[!h]
\centering
\includegraphics[height=7cm]{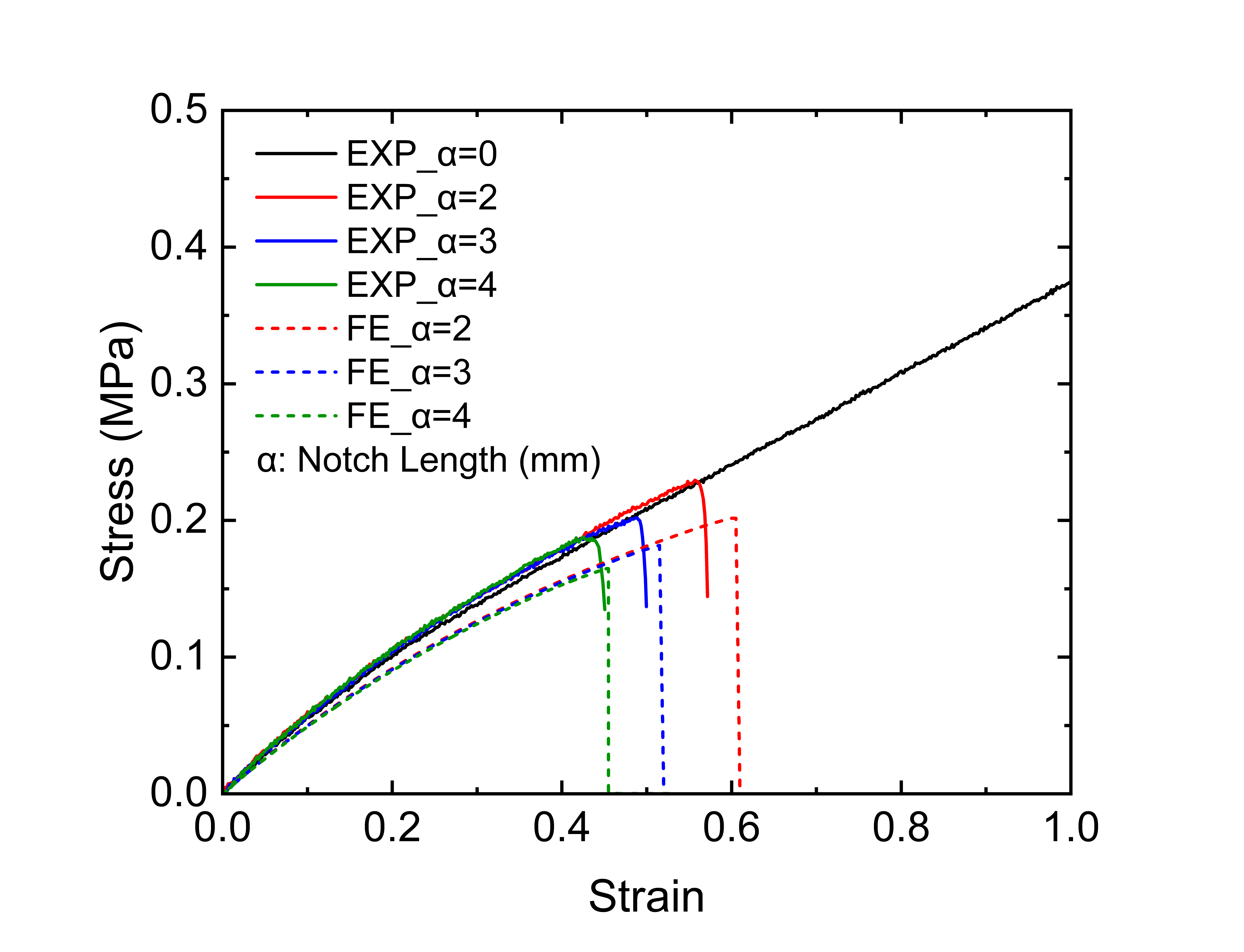}
\caption{Comparison between the numerical and experimental stress-strain curves for samples with different notch lengths (\(\alpha = 2, 3, 4\)mm) during SENT tests.}
\label{Fig5}
\end{figure}

The numerical and experimental stress-strain curves corresponding to the SENT tests are shown in Fig. \ref{Fig5}. It can be observed that the experimental curves exhibited a high degree of similarity before the samples failed. This allows for the determination of the material constant \(C_1\), which was found to be 0.09 MPa based on the experimental data prior to failure. Additionally, the critical strain energy release rate \(G_c\) was calculated to be 0.27 N/mm using eqns (\ref{5.36}) and (\ref{5.37}). Subsequently, the values of \(C_1\) and \(G_c\) were employed in the FE simulations to calibrate the length scale \(l\) in the phase field method. The results showed that when \(l = 1\) mm, the experimental and simulated results agreed well for all different notch length tests (Fig. \ref{Fig5}).

\begin{figure*}
 \centering
 \includegraphics[height=8cm]{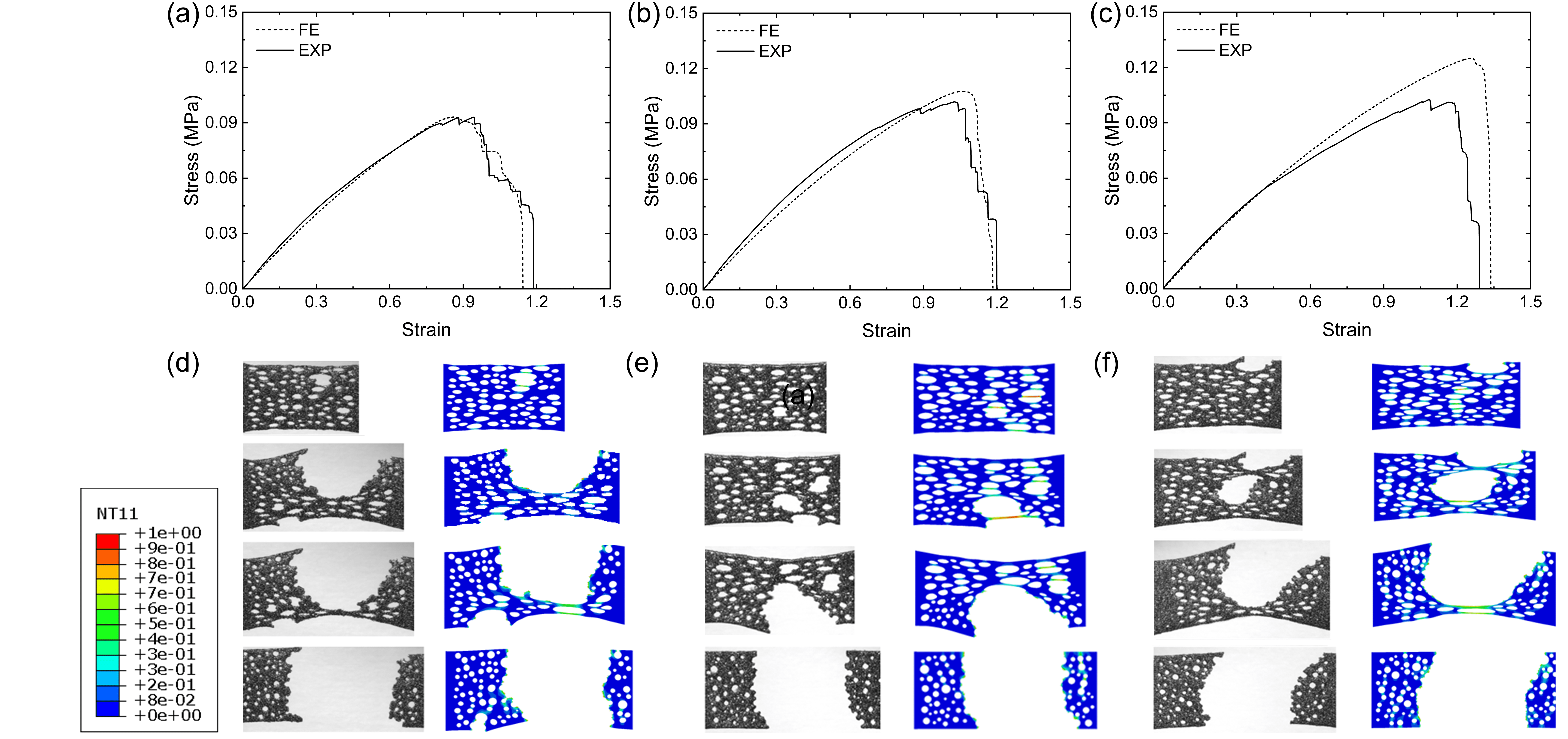}
 \caption{Comparison between the (a-c) numerical and experimental stress-strain curves and (d-f) fracture paths for three porous samples in tension (Elements with \(NT11\geq1\) were deleted).}
 \label{Fig6}
\end{figure*}

In addition, three porous samples with the same scaled distribution parameters and porosity (Fig. \ref{Fig6}) were 3D printed to validate the rate-dependent phase field model and to calibrate the fracture parameter \(\eta\). The results showed that when \(\eta=0.4\), both the stress-strain curves and the failure paths of the simulated and experimental results were in good agreement for all three porous structures (Fig. \ref{Fig6}). Therefore, the material parameters for TangoBlack Plus FLX980 used in the phase field method are summarised in Table \ref{tbl:5.2}.

\begin{table}[h]
\small
  \caption{\ Material parameters for TangoBlack Plus FLX980.}
  \label{tbl:5.2}
  \begin{tabular*}{0.48\textwidth}{@{\extracolsep{\fill}}ll}
    \hline
    Material parameters & values  \\
    \hline
    \(k\) & 1   \\ 
    \(C_1 (MPa)\) & 0.09  \\ 
    \(G_c (N/mm)\) & 0.27   \\ 
    \(l (mm)\) & 1   \\ 
    \(\eta\) & 0.4   \\ 
    \hline
  \end{tabular*}
\end{table}

\section{Results of the RVEs analysis}
\subsection{The effect of large-scale pore distribution on mechanical behaviour}
\begin{figure*}
 \centering
 \includegraphics[height=11cm]{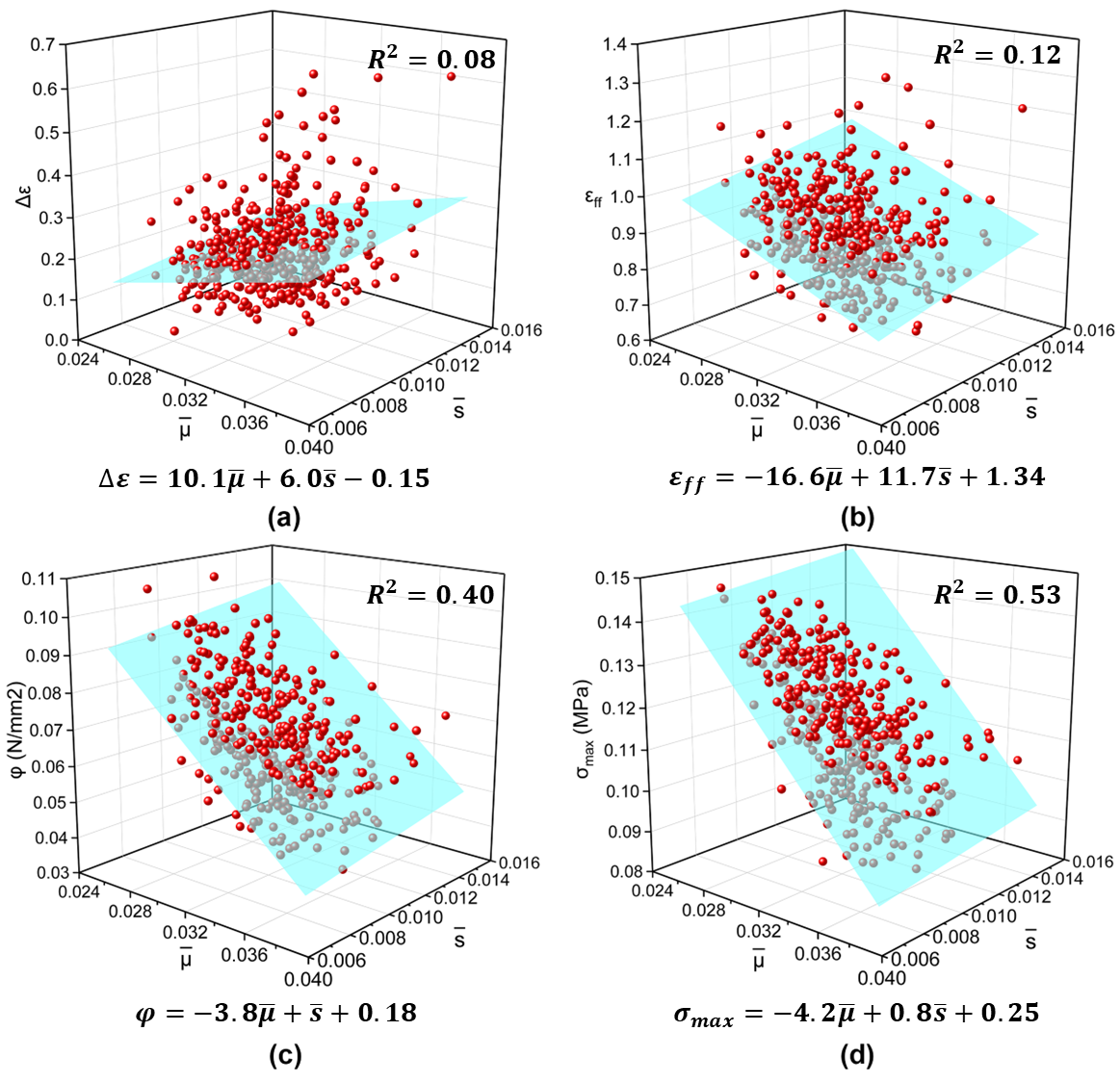}
 \caption{The RVE simulation and 3D linear regression results between the normalised mean value standard $\bar{\mu}$, normalised deviation value $\bar{s}$ of the large-scale pore radius and (a) \(\Delta\varepsilon\) with a coefficient of determination ($R^2$) value of 0.08, (b) final failure strain (\(\varepsilon_{f f}\)) with a $R^2$ value of 0.12, (c) strength (\(\sigma_{max}\)) with a R² value of 0.4, and (d) strain energy density (\(\varphi\)) with a $R^2$ value of 0.53. The blue planes represent the linear regression surfaces, and the red dots represent the RVE results.}
 \label{Fig7}
\end{figure*}

The normalised mean ($\bar{\mu}$) and standard deviation ($\bar{s}$) of pore radius reflect the statistic characteristic of the pore distribution within the single-scale RVEs, raising the question of whether these two parameters have a dominant effect on mechanical behaviour. To address this, Fig. \ref{Fig7} presents the simulation results for 400 samples, showing the difference between the initial and final failure strain (\(\Delta\varepsilon\)), final failure strain (\(\varepsilon_{f f}\)), strength (\(\sigma_{max}\)), and strain energy density (\(\varphi\)) as functions of the normalised mean ($\bar{\mu}$) and standard deviation ($\bar{s}$). The range of $\bar{\mu}$ (0.025-0.038) and $\bar{s}$ (0.007-0.015) of the RVE results shown in Fig. \ref{Fig7} closely mirrors the range of \(\bar{\mu}_p^*\) and \(\bar{s}_p^*\) observed in the SEM images (Figs. \ref{Fig1}e, \ref{Fig2}, S1 and S2 in the ESI,\dag~), indicating that the RVE results could reflect the geometrical effect of the microstructures on the mussel plaques. 

3D linear regression analysis was used as an estimation method to evaluate the trend of data as functions of $\bar{\mu}$ and $\bar{s}$ (Fig. \ref{Fig7}) \cite{freedman2009statistical}:
\begin{equation}
\hat{o}=a_0+a_1 \bar{\mu}+a_2 \bar{s}
\end{equation}
where \(\hat{o}\) represents the predicted value from the linear regression, and \(a_0\), \(a_1\), \(a_2\) are the regression coefficients. 

The Coefficient of Determination \(R^2\) was calculated to measure how well the regression model predicts the real data points (Fig. \ref{Fig7}) \cite{devore2000probability}:
\begin{equation}
R^2=1-\frac{\sum_{i=1}^{\xi}\left(o_i-\hat{o}_i\right)^2}{\sum_{i=1}^{\xi}\left(o_i-\bar{o}\right)^2}
\end{equation}
where \(o_i\) is the i th observed value (i.e., \(\Delta\varepsilon\), \(\varepsilon_{f f}\), \(\sigma_{max}\), and \(\varphi\)), \(\hat{o}_i\) is the predicted value, \(\bar{o}\) is the mean of observed values and \(\xi\) is the number of data points (i.e., \(\xi\)=400). 

\(R^2\) ranges from 0 to 1, where 1 indicates perfect fit and 0 indicates that the regression model cannot account for any of the variation in the response data around its mean \cite{nagelkerke1991note}. The nonlinear regression analysis was also conducted to fit the data. However, there was little improvement of the value of \(R^2\). For simplicity, only results of the linear regression analysis are presented in Fig. \ref{Fig7}.

\begin{figure*}
 \centering
 \includegraphics[height=10cm]{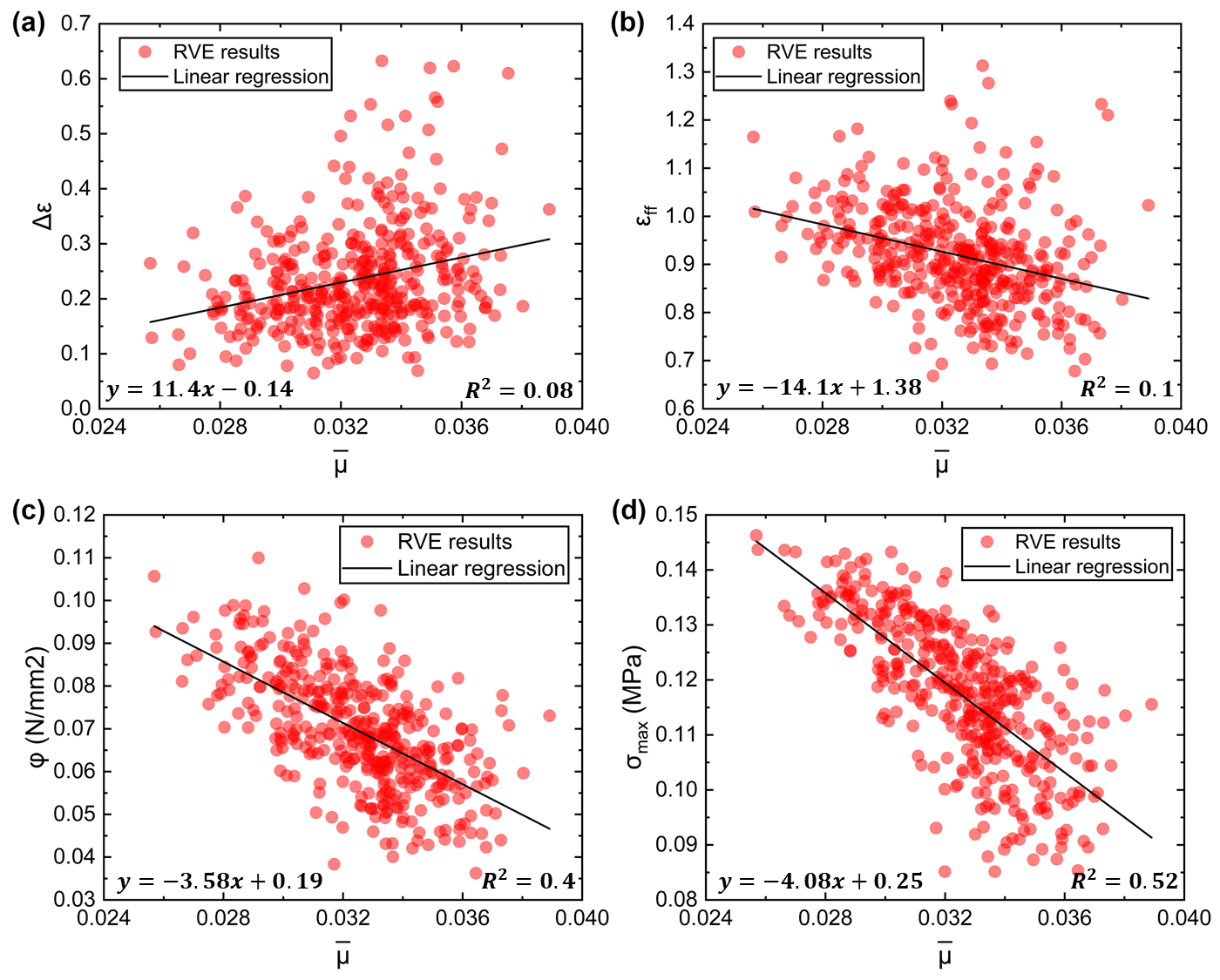}
 \caption{The 2D linear regression analysis between the normalised mean value standard $\bar{\mu}$ of the large-scale pore radius and (a) \(\Delta\varepsilon\), (b) final failure strain (\(\varepsilon_{f f}\)), (c) strength (\(\sigma_{max}\)), and (d) strain energy density (\(\varphi\)).}
 \label{Fig8}
\end{figure*}

The results of the linear regression, as shown in Fig. \ref{Fig7}, demonstrate that the overall effects of the normalised standard deviation ($\bar{s}$) on \(\Delta\varepsilon\), \(\varepsilon_{f f}\), \(\sigma_{max}\), and \(\varphi\) were statistically insignificant for the 400 samples. Conversely, the regression analysis indicates that an increase in the normalised mean radius ($\bar{\mu}$) led to an overall decreasing trend in strength (\(\sigma_{max}\)), strain energy density (\(\varphi\)), and final failure strain (\(\varepsilon_{f f}\)), respectively (Figs. \ref{Fig8}b, c and d). While a different overall trend was observed for \(\Delta\varepsilon\): as $\bar{\mu}$ increased, the value of \(\Delta\varepsilon\) exhibited a slight increase (Fig. \ref{Fig8}a). 

It is important to note that linear regression analysis only illustrates the overall trend of the effect of the normalised mean ($\bar{\mu}$) and standard deviation ($\bar{s}$) on macroscopic material properties and cannot capture the high dispersion of the results (Figs. \ref{Fig7} and \ref{Fig8}). It can be seen from Fig. \ref{Fig8} that the same $\bar{\mu}$ can always return scattered \(\Delta\varepsilon\), \(\varepsilon_{f f}\), \(\sigma_{max}\), or \(\varphi\) values. This may suggest that, while the normalised mean value ($\bar{\mu}$) of pore radius can influence the ductility or other macroscopic material properties of the single-scale RVE, it might not be the sole determining factor. Other factors such as the interactions between pores or the multi-scale structure could also play an important role. 

\subsection{Sudden and progressive failure modes within the single-scale RVEs}
\label{singleRVE}
\begin{figure}[h]
\centering
  \includegraphics[height=11cm]{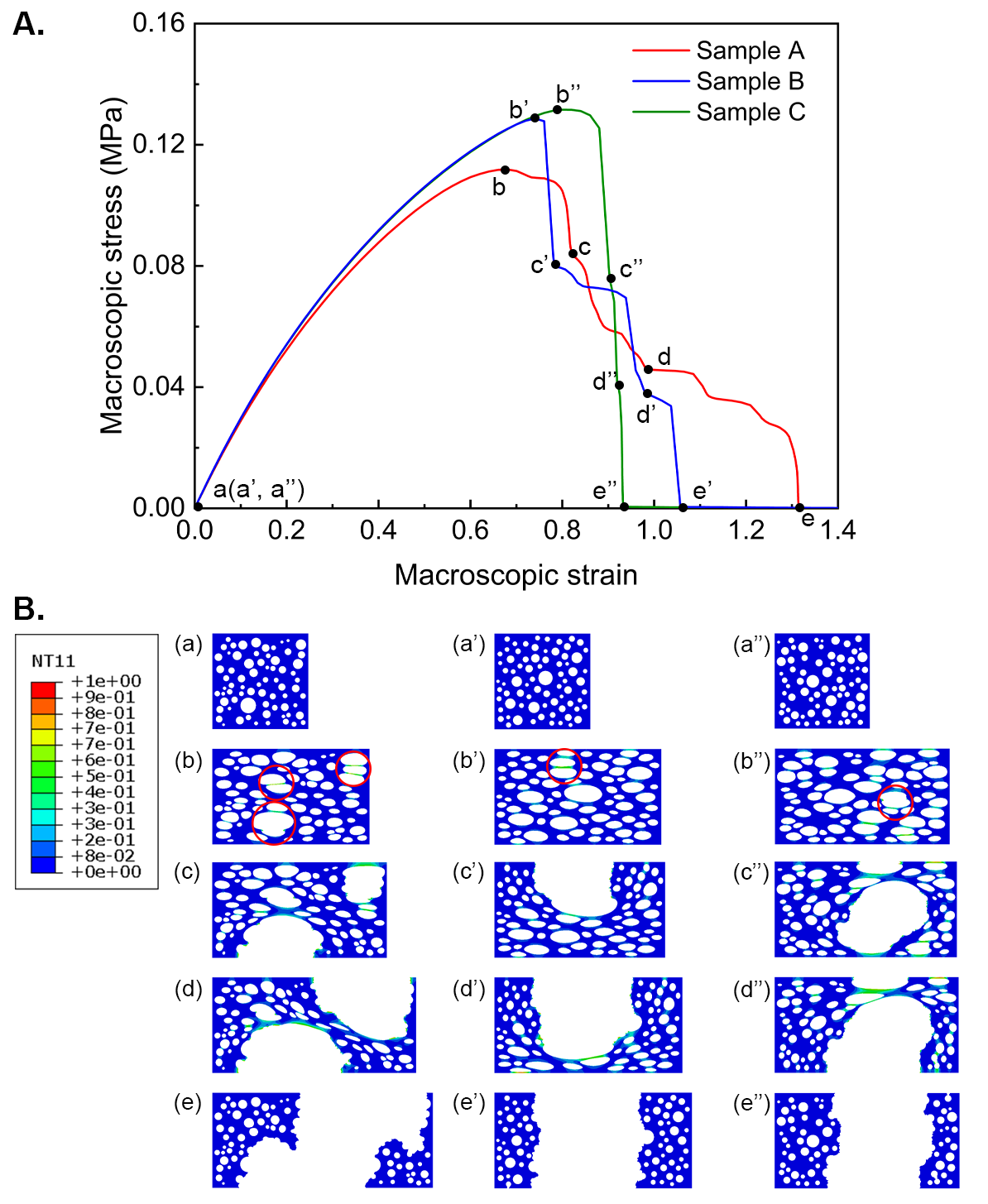}
  \caption{(A) The macroscopic stress-strain curves, and (B) crack paths of selected samples with different failure modes.}
  \label{Fig9}
\end{figure}

To better understand the failure mechanism of the porous RVEs, Fig. \ref{Fig9} shows the macroscopic stress-strain curve and crack propagation of Sample A, which had the highest values of \(\Delta\varepsilon\), \(\varepsilon_{f f}\), and \(\varphi\) among the 400 RVE results. For comparison, the stress-strain curves and crack propagation of two additional samples, B and C, are also presented in Fig. \ref{Fig9}. It is noted that samples A, B and C have nearly identical normalised standard deviation ($\bar{s}$) and mean ($\bar{\mu}$) values, with a difference of only ±0.001. 

Two different failure modes were identified from the stress-strain curves of the samples: progressive failure and sudden failure (Fig. \ref{Fig9}A). For the progressive failure sample (Sample A), the \(\Delta\varepsilon\) value was approximately 50\% of the final failure strain. In contrast, the \(\Delta\varepsilon\) value for the sudden failure sample (Sample C) was close to 0 (Fig. \ref{Fig9}A). The final failure strain of the progressive failure sample (Sample A) was about 1.5 times that of the sudden failure sample (Sample C), but its strength was only approximately 85\% of Sample C’s strength (Fig. \ref{Fig9}A). Interestingly, the strength of these samples decreased with increasing \(\Delta\varepsilon\) (Fig. \ref{Fig9}A), indicating that the high ductility of the plaques may have been achieved by sacrificing strength. 

The crack propagation of Samples A, B and C are presented in Fig. \ref{Fig9}B to further investigate their failure mechanisms. It can be seen from Fig. \ref{Fig9}B that although the normalised standard deviation ($\bar{s}$) and mean ($\bar{\mu}$) of Samples A, B and C are nearly identical, their underlying failure mechanisms are distinct. The damage observed in Sample A initiated at three distinct locations: large pores or closely spaced pores aligned perpendicular to the tensile direction (Fig. \ref{Fig9}B). The scattered locations of the initial damage caused a diffused crack pattern with elongated fracture path, resulting in the progressive failure mode of Sample A. In Sample C, the initial damage occurred at closely clustered pores perpendicular to the tensile direction (Fig. \ref{Fig9}B), rapidly forming a complete fracture path, i.e. concentrated crack pattern. Although another damage was also observed during the failure process, it played a minor role as the initial damage was the primary cause of the failure (Fig. \ref{Fig9}B). 

Sample B exhibited a fracture path similar to Sample C (Fig. \ref{Fig9}B). However, unlike the rapidly developing failure observed in Sample C, the failure in Sample B was more progressive (Fig. \ref{Fig9}A). It might be because the blunted cracking observed in Sample B enabled the stress to be distributed and absorbed more energy. While the pore distribution in Sample C may cause the blunted cracking to be insufficient to prevent the rapid crack growth and stress concentration. In addition to the distribution of large-scale pores, the interactions among pores might also influence the failure mechanisms of the RVEs.

\subsection{The effect of multi-scale structures on mechanical behaviour}
\begin{figure}[h]
\centering
  \includegraphics[height=10cm]{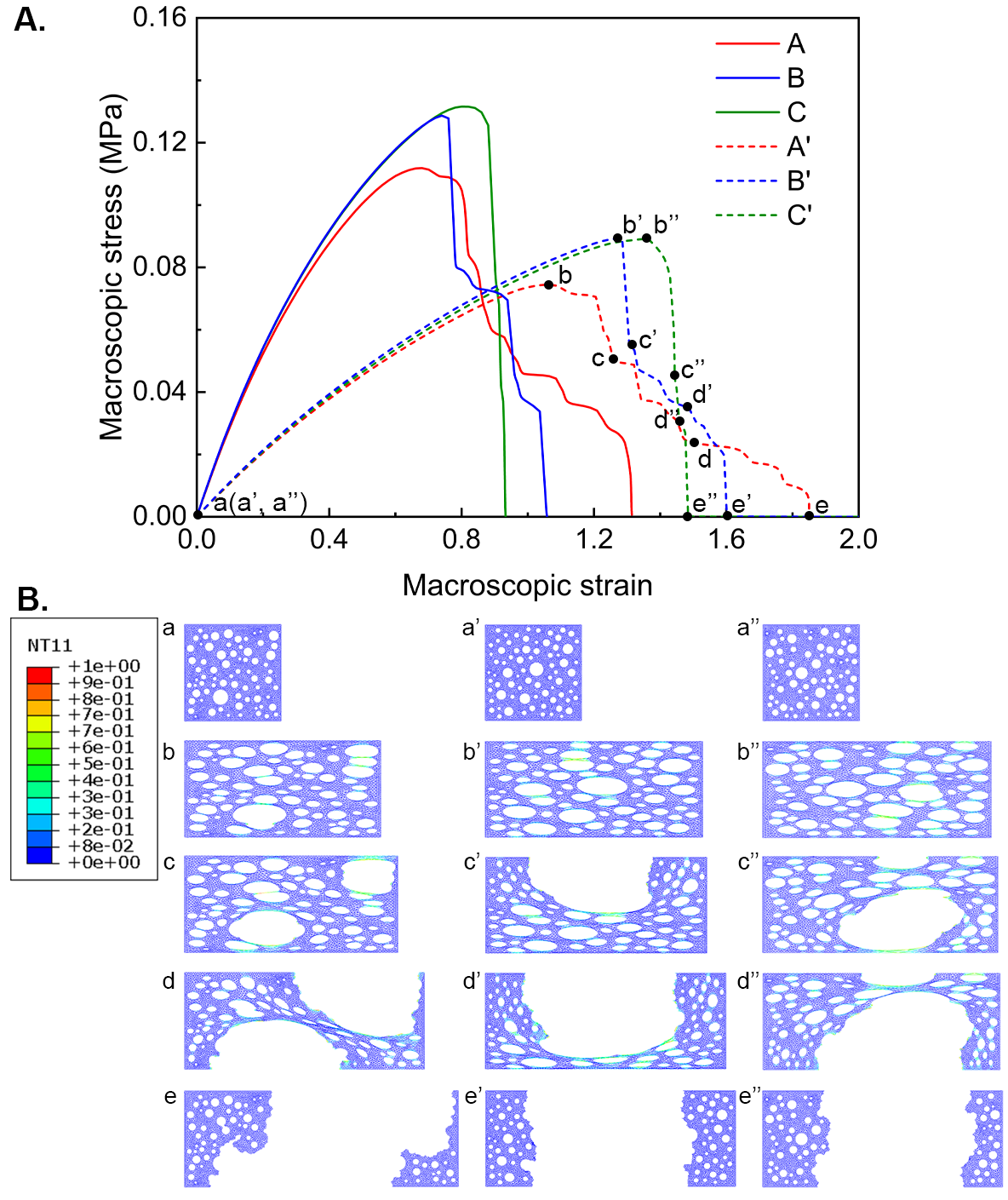}
  \caption{(A) The macroscopic stress-strain curves of single and multi-scale porous samples, and (B) crack paths of multi-scale porous samples.}
  \label{Fig10}
\end{figure}

The SEM images of mussel plaque core revealed the presence of two different scales of pores (Fig. \ref{Fig1}e). In addition to the large-scale pores that were analysed above, small-scale pores were found to fill the remaining areas of mussel plaques. New multi-scale porous RVEs based on the three samples selected in Section \ref{singleRVE} were regenerated and analysed to understand the effect of the multi-scale porous structure on the failure mechanism (Fig. \ref{Fig10}). 

A comparison of the fracture paths shown in Figs. \ref{Fig9}B and \ref{Fig10}B reveals that the failure modes were almost identical for all three samples, despite the structural change from the single-scale porous RVE to the multi-scale porous RVE. This suggests that the failure mechanism could be dominated by the large-scale pores.  

The stress-strain curves for both multi-scale and single-scale RVEs are presented in Fig. \ref{Fig10}A. The multi-scale porous RVEs with 32\% porosity exhibited significantly lower stiffness compared to the single-scale RVEs with 27\% porosity, despite having only a 5\% difference in porosity (Fig. \ref{Fig10}A). To understand this mechanism, numerical simulations of solid RVE and small-scale porous RVE with 5\% porosity were conducted. These simulations demonstrate that even a small increase in porosity can lead to a substantial decrease in material stiffness due to microstructural changes (Fig. S8 in the ESI,\dag~). The reduction in stiffness, however, enhanced the ductility of the multi-scale porous RVEs (Fig. \ref{Fig10}A). Specifically, the final failure strains of the three multi-scale samples increased by 40\% (Sample A’), 50\% (Sample B’), and 60\% (Sample C’), respectively, compared to the single-scale samples (Fig. \ref{Fig10}A). Meanwhile, the strength of the three multi-scale samples decreased by 33\% (Sample A’), 31\% (Sample B’), and 32\% (Sample C’), respectively, due to the reduction in stiffness. This may suggest that the multi-scale porous structure of mussel plaques might improve the ductility by reducing their stiffness, but this could also result in a reduction in strength. 

\section{Discussion}
Researchers have consistently focused on understanding the structural and chemical compositions of mussel plaques and their impact on adhesion \cite{burkett2009method,desmond2015dynamics,waite2017mussel,filippidi2015microscopic}. Previous study has suggested that the pore arrangement within the plaque core might influence the adhesion ability \cite{ghareeb2018role}. However, limited attention has been paid to understand how pore arrangement affects the mechanical behaviours of the plaque core. Tensile testing of the thread-plaque system reveals that mussel plaques can withstand considerable deformation before detachment, underscoring their high ductility (Fig. \ref{Fig1}c) \cite{pang2023quasi,lyu2024determining}. This led us to hypothesise that the pore distribution and hierarchical structure of the plaque core might also contribute to its high ductility and potentially impact other tensile properties. 

Elbanna et al. (2015) conducted a semi-analytical approach to understand the effect of void arrangement on the adhesion of mussel plaques \cite{ghareeb2018role}. They found that plates with larger voids near the attached interface display higher stress heterogeneities, which can trap cracks in regions with low stress, thereby slowing down the propagation of cracks and resulting in higher maximum forces, increased ductility, and enhanced energy dissipation. However, the ordered and graded pores were used in the study, which differ from the natural pore distribution found in the plaque core and may influence the mechanical behaviour \cite{ghareeb2018role}. 

In this research, extensive SEM analysis was employed to obtain the pore distribution of the plaque core, generating porous RVEs using scaled distribution parameters to present the main geometric characteristics of the plaque core. Our results demonstrate that large-scale pore distribution significantly impacts the mechanical behaviour of single-scale porous RVEs, including ductility, strength, and strain energy density. 

As mentioned above, Elbanna et al. (2015) reached the conclusion that larger voids near the attached interface can lead to higher strength, ductility, and strain energy density \cite{ghareeb2018role}. Our RVE-based numerical studies have suggested that increasing the normalised mean value ($\bar{\mu}$) of pore radius can statistically lead to a decreasing trend in these macroscopic material properties: ductility, strength, and strain energy density. While the normalized standard deviation value ($\bar{s}$) of pore radius had an insignificant effect on these properties. Additionally, it was found that the same normalised mean value ($\bar{\mu}$) or normalized standard deviation value ($\bar{s}$)  can always return highly scattered macroscopic material property values. This indicates that the values of $\bar{\mu}$ and $\bar{s}$ can not solely determine the mechanical behaviour of single-scale porous RVEs. Other factors such as the interactions between pores and the hierarchical structure could also play an important role.

Regarding to the interactions between pores, our FE simulation results suggest that two types of failure mechanisms might occur under the same pore distribution, i.e. diffused crack pattern and concentrated crack pattern. The diffuse crack pattern elongated the crack path, which in turn led to the progressive failure mode and an increase in the ductility of RVEs. While the concentrated crack pattern led to the rapid crack propagation and resulted in a sudden failure mode. The observed findings highlight the crucial role played by pore interactions in influencing the mechanical behaviours. 

Another factor, i.e. the hierarchical structure, was also examined in this research. The results indicates that the failure modes of the porous RVEs are dominated by the large-scale pores, but the hierarchical structure can further increase the ductility of porous RVEs by reducing stiffness, suggesting it is also a crucial factor to achieve the high ductility of mussel plaques. 

This study advances the understanding of how pore distribution and hierarchical structure within the mussel plaque core contribute to its high ductility and other material properties. These insights are crucial for mimicking the unique microstructure in the design of advanced materials. For instance, Bosnjak and Silberstein (2021) presented two chemical pathways to develop ductile materials with high toughness. Our research shows that altering the pore distribution or introducing the hierarchical structure might be another potential pathway to design materials with combined strong properties \cite{bosnjak2021pathways}. However, several limitations remain. For instance, the simplified 2D structural analysis cannot fully capture the complexity of actual 3D structures. The multi-scale porous RVEs are an approximation rather than an exact replication of the multiscale porous structure of mussel plaques. Future research should focus on extending this study to the three-dimensional level, as well as further accurate modelling of the microstructure of mussel plaques. 

\section{Conclusions}
This study has systematically explored the influence of the pore distribution and hierarchical porous structures within the mussel plaque core on its tensile properties, with a specific emphasis on enhancing ductility. Our main conclusions are summarized as follows:

\begin{enumerate}
    \item The SEM analysis of mussel plaque core shows there are two different scale of pores within the plaque core: small-scale pores with diameters ranging from 50 nm to 800 nm and large-scale pores with diameters ranging from 1 \(\mu m\) to 3 \(\mu m\). The size of large-scale pores follows a lognormal distribution, while the location of them follows a uniform distribution. 
    \item The large-scale pore distribution can significantly influence the ductility, strength, and strain energy density of the porous RVEs. Statistically, increasing the normalised mean value ($\bar{\mu}$) of pore radius can lead to a decreasing trend in the value of the ductility, strength and strain energy density. However, the normalised standard deviation value ($\bar{s}$) of pore radius had an insignificant effect on these properties. Additionally, the same values of $\bar{\mu}$ or $\bar{s}$ can always return scattered values of the ductility, strength and strain energy density, indicating that they cannot solely determine the macroscopic material properties.
    \item Two distinct failure modes (sudden and progressive) were identified under the same pore distribution. The interactions between pores led two different crack patterns: diffused and concentrated. The diffused crack pattern extended the crack path and delayed the failure, which led to a progressive failure mode and higher ductility. While the concentrated crack pattern had rapid crack propagation, resulting in a sudden failure mode but with higher strength. This indicates that mussels may enhance the ductility of their plaques by sacrificing plaques’ strength to form the porous plaque core.
    \item The multi-scale porous RVEs exhibited identical failure modes to the single-scale porous RVEs, indicating that the failure mechanism of the mussel plaque could be dominated by the large-scale pores. Compared to the single-scale porous RVEs, multi-scale truss RVEs had 40\%-60\% higher ductility but lower stiffness and strength. This suggests that the hierarchical structure of mussel plaques might enhance their ductility by reducing the stiffness. 
\end{enumerate}

\section*{Author contributions}
Y.L.: conceptualization, data curation, formal analysis, investigation, methodology, software, validation, visualization, writing—original draft, writing—review and editing;
M.T.: conceptualization, investigation, methodology, software, writing—review and editing; Y.P.: formal analysis, investigation, methodology, validation, writing—review and editing; W.S.: supervision, writing—review and editing; S.L: supervision, writing—review and editing; T.L.: conceptualization, data curation, funding acquisition, investigation, methodology, project administration, supervision, validation, writing—original draft, writing—review and editing.

\section*{Data availability}
The data supporting this article have been included as part of the Supplementary Information.

\section*{Conflicts of interest}
There are no conflicts to declare.

\section*{Acknowledgements}
The authors thank the support provided by the Leverhulme Trust Research Grant Scheme (RPG-2020-235), the NanoCAT Research Grant of the University of Nottingham, and the Centre for Additive Manufacturing of the University of Nottingham.

%
%
%


\balance


\bibliography{rsc} 
\bibliographystyle{rsc} 

\end{document}